\documentclass[10pt,twocolumn,journal]{IEEEtran}
\usepackage{amsmath,amssymb,amsfonts}
\usepackage{graphicx}
\usepackage{textcomp}
\usepackage{xcolor}
\usepackage{balance}
\usepackage[normalem]{ulem}
\usepackage{cite}
\usepackage{url}
\usepackage{subfig}
\usepackage{booktabs}

\begin{document}

\title{ISAC-Enabled On-Demand UAV Charging for Wireless Rechargeable Sensor Networks}


\author{Muhammad Umar Farooq Qaisar, Lin Zhang, Paolo Bellavista, Shehzad Ashraf Chaudhry, Shamsher Ullah, and Chang Liu
	\thanks{M. U. F. Qaisar (muhammad@buaa.edu.cn) is with the Hangzhou International Innovation Institute of Beihang University, Hangzhou 311115, China}
	\thanks{L. Zhang (zhanglin@buaa.edu.cn) is with Hangzhou International Innovation Institute of Beihang University, Hangzhou 311115, China, the School of Automation Science and Electrical Engineering at Beihang University, Beijing 100191, China, and the State Key Laboratory of Intelligent Manufacturing Systems Technology, Beijing 100854, China}
	\thanks{P. Bellavista (paolo.bellavista@unibo.it) is with the Department of Computer Science and Engineering, University of Bologna, Bologna, Italy}
	\thanks {S. A. Chaudhry (shehzad.ashraf@adu.ac.ae) is with the Department of Computer Science and Information Technology, College of Engineering, Abu Dhabi University, Abu Dhabi, UAE; and with the Department of Software Engineering, Faculty of Engineering and Architecture, Nisantasi University, Istanbul, Turkey}
	\thanks{S. Ullah. (shamsher@mail.ustc.edu.cn) is with the School of Artificial Intelligence, Shenzhen University, Shenzhen, P.R China}
	\thanks{C. Liu (liuchang@gdut.edu.cn) is with the School of Information Engineering, Guangdong University of Technology, Guangzhou 510006, China}
	\thanks{Corresponding authors: Muhammad Umar Farooq Qaisar, Email: muhammad@buaa.edu.cn}}

\maketitle

\begin{abstract}
Unmanned aerial vehicles (UAVs) equipped with wireless power transfer (WPT) extend the lifetime of wireless rechargeable sensor networks (WRSNs) by delivering energy on demand.  This article presents an integrated sensing and communication (ISAC)-enabled on-demand UAV charging framework coordinated by a central base station. A prioritized charging queue captures node urgency and service cost through residual energy, traffic load, estimated UAV travel time, and flight-direction alignment. This bidirectional coupling ensures that scheduling decisions shape the UAV trajectory, while updated mobility estimates from ISAC dynamically reorder the queue. ISAC-assisted estimation of UAV distance, speed, and position updates travel-time predictions under mobility uncertainty. A time-allocated partial charging policy distributes limited hover time across queued nodes according to criticality. Simulations show gains in energy usage efficiency, travel distance, and charging delay compared
with representative baselines. We discuss deployment considerations, including computational overhead, scalability, and parameter selection, to aid practitioners evaluating the framework for IoT scenarios.
\end{abstract}

\begin{IEEEkeywords}
Wireless rechargeable sensor networks, UAV charging, wireless power
transfer, integrated sensing and communication, on-demand scheduling,
partial charging.
\end{IEEEkeywords}

\section{Introduction}

Wireless sensor networks (WSNs) form the core sensing substrate for the Internet of Things (IoT), enabling continuous monitoring in infrastructure health, precision agriculture, industrial automation, and environmental surveillance.  Despite advances in low-power hardware and energy-aware protocol design, most deployments remain fundamentally energy-limited because sensing, computation, and multi-hop communications draw from finite on-board batteries \cite{1.1}.  Energy depletion at a small subset of nodes can have a disproportionate impact: a failed relay may disconnect an entire region, while the loss of critical sensing points reduces coverage and degrades inference quality \cite{1.2}.  These failure modes motivate wireless rechargeable sensor networks (WRSNs), where wireless power transfer (WPT) replenishes node energy and enables long-term operation \cite{1.3}.

\begin{figure}[h]
\centering
\includegraphics[width=\columnwidth]{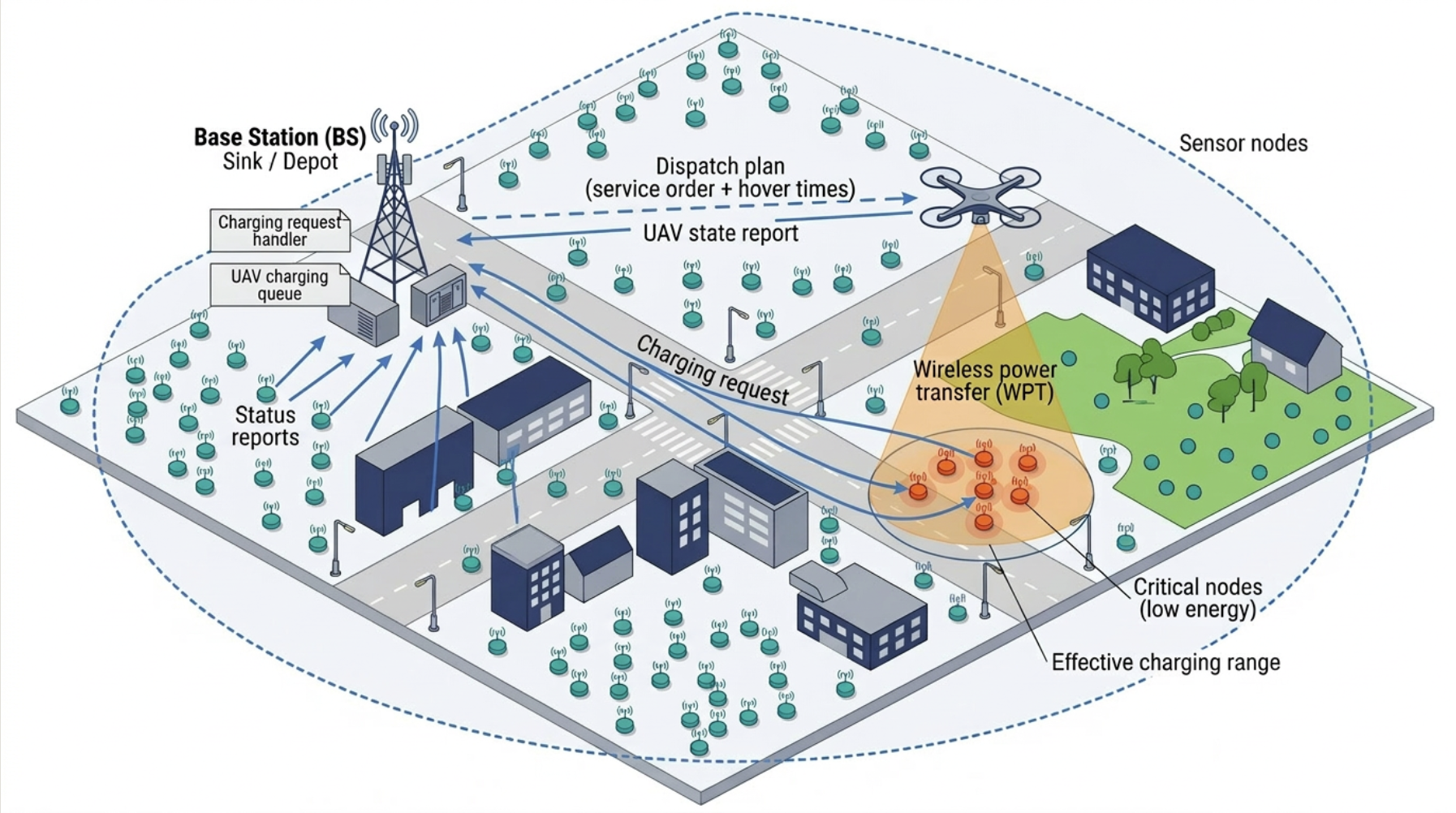}
\caption{UAV-assisted wireless rechargeable sensor network (WRSN) scenario: the base station (BS) dispatches a WPT-equipped UAV to hover and recharge sensor nodes within an effective charging range.}
\label{fig:network}
\end{figure}

In WRSNs the primary bottleneck shifts from node energy capacity to charging scheduling and route planning under limited mission time and energy. A mobile charger has to perform three tasks: which nodes to serve, in what order, and for what duration, while working online as requests keep coming in. The scheduler tries to balance urgency (avoiding outages) with service cost (the time and energy to get to and charge a node), and it must do so online as requests come in \cite{1}.  The challenges amplified by using unmanned aerial vehicles (UAVs) as mobile chargers are: UAVs are able to bypass the obstacles on the ground. They would reach the dispersed nodes quickly. Nevertheless, the limited flight endurance and expensive hovering tightly couples routing and scheduling. Every additional detour increases propulsion expenditure besides causing a rise in the mission time. This reduces the charging time available for other nodes. Besides, it raises the risk of charging deadlines being missed \cite{1.5}. Fig.~\ref{fig:network} illustrates the considered UAV-assisted WRSN scenario, where a base station coordinates a WPT-equipped UAV that travels to low-energy sensors and hovers to recharge them.

A further practical challenge is that effective scheduling depends on accurate, timely knowledge of the UAV state and the evolving service cost to each candidate node. Travel time is not static, it varies with the UAV's current position, instantaneous speed, and the geometry of the partially executed tour. Consequently, route-planning and scheduling strategies based on low-fidelity or delayed mobility state estimates may incur route oscillations, unnecessary trajectory adjustments, and degraded prioritization.  This motivates a sensing-assisted control loop in which UAV state estimation and queue management are coupled \cite{1.6}. In particular, integrated sensing and communication (ISAC) \cite{1.4} enables the network to jointly support data exchange and obtain updated estimates of UAV distance, speed, and position \cite{1.7}. These estimates are used to refine access-time predictions and charging priorities, improving responsiveness and trajectory efficiency \cite{1.8}

This article develops OD-UCS, an ISAC-enabled on-demand UAV charging framework for WRSNs coordinated by a central base station. The framework helps in enhancing the decision quality in uncertain mobilities in constrained mission time.  The design is distinct for 3 features. A queuing mechanism for the charging of nodes is built that gives priority to service cost and node urgency using residual energy, traffic load, estimated UAV travel time, and compatibility with flight direction.  Moreover, It uses the ISAC-assisted estimation of UAV distance, speed, and position to continuously update travel-time predictions on the fly to enhance the reliability of scheduling decisions as UAV moves. Third, it introduces a time allocation-based partial charging policy, which distributes a limited hover time among the queued nodes based on their criticality to cover more nodes while missing fewer deadlines. Since the queue explicitly considers travel time and alignment of the travelling direction, both determined by the UAV’s current position and heading, any scheduling decision implicitly shapes the trajectory, and vice versa, trajectory updates prompted via ISAC estimation trigger a reordering of priority. The design contains a close bidirectional coupling.

The rest of this article is structured as follows. Section~\ref{sec:related} presents a review of related work. It positions our contribution with respect to the literature. Section~\ref{sec:system} presents the system model encompassing the network, charging and mobility components. The proposed charging model with priority queue, ISAC-aided state estimation, time-allocated partial charging is detailed in Section~\ref{sec:proposed}. Section~\ref{sec:performance} evaluates and discusses the performance. The article is concluded in Section~\ref{sec:conclusion} and future research directions are outlined.

\section{Related Work} \label{sec:related}

UAV-enabled WRSNs have evolved from basic tour-planning formulations to integrated service frameworks that must balance mobility limits, heterogeneous node depletion, and timely response under realistic deployment constraints. Rather than surveying each contribution in isolation, we organize the discussion along three design dimensions, architecture and coordination, trajectory and energy management, and charging scheduling intelligence, and identify where prior work leaves gaps that our framework addresses.

\textbf{Architecture and coordination.}
Early work focused on hybrid vehicle-drone architectures. Chen~\emph{et~al.}~\cite{1.5} proposed a collaborative model that combines wireless charging vehicles (WCVs), WCV-carried drones, and independent drones, using a ring-based partition to coordinate responsibilities and mitigate the respective limitations of ground vehicles and drones in large WRSNs. Zhao~\emph{et~al.}~\cite{1.6} investigated UAV dispatch planning for bridge-monitoring WRSNs, emphasizing feasible routing under obstacle constraints. These works demonstrate multi-platform coordination but do not exploit ISAC for real-time state estimation within the scheduling loop.

\textbf{Trajectory and energy management.}
Lin~\emph{et~al.}~\cite{1.8} formulated a period-area coverage perspective in which a UAV must both recharge sensors and provide periodic sensing coverage.  Chen~\emph{et~al.}~ \cite{2.1} presented CGDA-Q for agricultural IoT, combining adaptive charging cells and Q-learning for path planning.  Liu~\emph{et~al.}~\cite{2.4} studied an IRS-assisted UAV charging architecture with a two-stage trajectory and phase-optimization approach. Ma~\emph{et~al.}~\cite{2.5} explored laser-powered UAV far-field wireless charging coupled with data backhauling. These approaches advance physical-layer energy delivery but generally assume accurate UAV state and do not couple state estimation with online priority updates.

\textbf{Charging scheduling intelligence.}
Liu~\emph{et~al.}~\cite{2.2} adopted deep reinforcement learning (DQN) for UAV-assisted charging to reduce sensor downtime. Betalo~\emph{et al.}~\cite{2.3} used multi-agent deep reinforcement learning to coordinate charging and path planning across multiple UAVs. Gupta~\emph{et~al.}~\cite{2.6} modeled UAV-enabled charging under a non-cooperative pricing framework among charging providers. While these learning-based approaches improve adaptability, they require extensive training and offer limited interpretability.

\textbf{Positioning of this work.}
Prior UAV-enabled WRSN charging studies improve routing and scheduling under energy and time constraints, yet they generally assume accurate mobility information and do not tightly couple state estimation with online priority updates. They also lack principled, time-allocated partial-charging policies.  Our framework integrates ISAC-assisted estimation into the scheduling loop, uses transparent priority scheduling that needs no training, and applies urgency-weighted partial charging to improve responsiveness under constrained endurance.

\section{System Model} \label{sec:system}

\subsection{Network Model and Assumptions}

We consider a WRSN deployed over a $500 \times 500$\,m$^2$.  A set of rechargeable sensor devices is randomly scattered within the area. Each sensor node is equipped with sensing, processing, and communication capabilities, as well as a rechargeable battery.

The base station is located at the center of the monitoring area. It acts as the sink for data traffic, serves as the controller for charging operations and a depot for the UAV. It maintains state information about sensor nodes, including their estimated positions, residual energy levels, and traffic statistics. This information is obtained through periodic status reports and, when necessary, through on-demand queries.

An unmanned aerial vehicle acts as an aerial mobile charger. The UAV is equipped with a WPT module for wirelessly transferring energy to sensor nodes and with communication interfaces for exchanging control and telemetry data. A UAV departs from the base station with enough energy to serve a subset of nodes and safely return to the base. Its movement is limited by the maximum speed limit, acceleration limits and flight time limit. 

The energy consumption profile of sensor nodes involves sensing, local processing and communication. Their energy consumption depends on their sensing duty cycle, traffic generation rate and their forwarding role in the network topology. Nodes send their remaining energy and traffic load to base station. They can communicate through many hops if they are not in direct range.

When a node's residual energy falls below a predefined threshold (30\,\%), it generates a charging request that includes the node's identifier, its estimated position, current residual energy, and recent
traffic statistics. The request is forwarded to the base station, which maintains a global queue of nodes awaiting charging. Nodes that exhaust their battery before being served enter a dormant state, they resume operation once recharged in a subsequent mission. This dormancy mechanism bounds the impact of missed deadlines and motivates the urgency-weighted design of the priority queue.

The base station manages the flight and charging operation of the UAV. The base station calculates the charging schedule, including which nodes the UAV should visit, when to visit them, and how long to charge each node. It updates this schedule as new charging requests arrive, and ISAC provides status updates on the UAV. The base station ensures that the UAV has sufficient energy for a safe return to the depot.

\subsection{Charging Model}

The WPT system of the UAV operates when the UAV hovers above the sensor node in a particular proximity. It is assumed that as long as the UAV is within an acceptable vertical and horizontal distance from a node, a near-constant charging rate can be achieved via alignment and coupling conditions. This simplifies the charging model while capturing a key dependence on dwell time.

Each operation of charging consists of a travel phase and a dwell phase. In the travel phase, the UAV moves to the neighborhood of the target node from its present position. The energy used for travelling in UAV is speed dependent. It changes altitude and make some diversion in order to avoid obstacles. During the dwell phase, the UAV remains stationary over the node and employs WPT to transfer energy. Hovering uses a lot of energy, which needs to be factored into mission budget.

We assume that either the energy consumption profile of the UAV is known or can be accurately estimated from previous calibration. The profile is used by the base station to decide on the number of nodes to be served in a single mission and to save some energy for return to the depot. The outcome is an efficient charging time budget for every mission, representing the total dwell time available among nodes.

This article abstracts out all the physical-layer details into charging rate parameter and focusses on the scheduling framework for allocating charging time.

\subsection{UAV Mobility Model}

The UAV follows a controlled stop-and-charge operation coordinated by the base station (BS).  It departs from the BS, travels in straight-line point-to-point flight between scheduled service locations at a constant speed $v_{\max}$, and hovers to recharge the selected sensor node. The BS updates the service sequence online as new requests arrive.

Before moving to the next target, the UAV verifies that its remaining onboard energy can support the flight to the candidate node, the planned hovering and charging duration at that node, and the	subsequent return trip to the base station including a safety margin.	\textbf{Return-to-depot safety constraint.}  The UAV may proceed to a candidate node only if its remaining energy exceeds the total of the propulsion energy needed to reach that node, the hovering energy	consumed during the allocated charging time at that node, and the energy required to fly back to the base station together with a safety	reserve. This constraint is applied before each node is admitted into the active charging queue, so that the return-to-depot guarantee is integrated into the scheduling policy rather than checked only at the end of a tour. If the remaining energy is insufficient, the node is deferred to the next mission and the UAV returns to the base station for energy replenishment.

The mobility constraints, maximum speed, endurance limit, and the return-to-depot guarantee, directly shape the scheduling decisions in Section~IV.  Because the priority queue accounts for estimated travel time and direction alignment (both functions of the UAV's current state), every reordering of the queue implicitly adjusts the trajectory. Conversely, when ISAC updates revise the UAV's estimated position or speed, the travel-time and direction terms in the priority formula change, which may trigger a different node ordering and hence a different planned path. This bidirectional coupling between scheduling and trajectory is a central design feature: it ensures that the UAV	follows a route that is both urgency aware and mobility efficient, adapting in real time as new information arrives.

\section{Proposed Charging Model} \label{sec:proposed}

\begin{figure*}[!t]
\centering
\includegraphics[width=\textwidth]{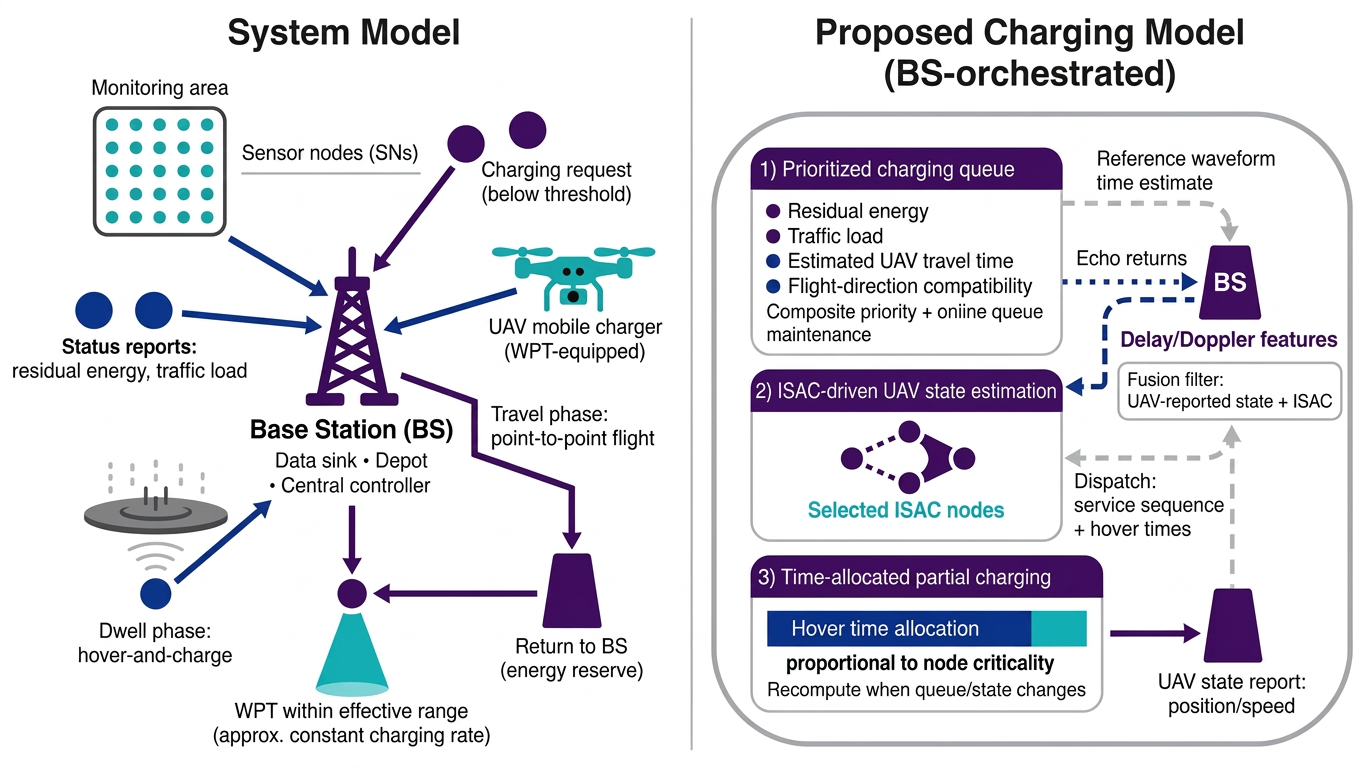}
\caption{Overview of the BS-orchestrated ISAC-enabled on-demand UAV charging framework for WRSNs: (left) system model showing status reporting, charging requests, and the UAV stop-and-charge operation with WPT within an effective range; (right) proposed charging model integrating (1) a prioritized charging queue, (2) ISAC-driven UAV state estimation with BS-side fusion, and (3) time-allocated partial charging with feedback from UAV state reports.}
\label{fig:overview1}
\end{figure*}

The proposed ISAC-enabled charging framework consists of three interrelated components: an ISAC-assisted prioritized charging queue, an ISAC-driven UAV state estimation mechanism, and a time-allocated partial charging model. The base station orchestrates these components to plan and continuously refine UAV missions in response to evolving network conditions and updated mobility estimates. Fig.~\ref{fig:overview1} provides a high-level view of the system model and the closed-loop interaction among queue-based scheduling, ISAC-assisted UAV state estimation, and time-allocated partial charging.

\subsection{ISAC-Assisted Prioritized Charging Queue}

The charging queue is designed for demand and cost representing the urgency of charging and costs of serving. The base station maintains four key attributes for every node in queue, namely residual energy, traffic load, estimated UAV travel time and direction alignment.

\subsubsection{Residual energy}
Residual energy is the most direct representation of charging urgency, as it quantifies how close a sensor is to energy depletion. Nodes that have extremely low remaining energy will fail before the UAV can get there. This leads to a deterioration in sensing coverage and, in multi-hop topologies, connectivity loss.

The base station also monitors the depletion rate of energy of each node. If one of them faces elevated consumption, their priorities may differ even if they have equal residual energy. As residual energy decreases, the urgency score increases, with a steeper response below a safety threshold. Consequently, the near-critical nodes move to the head of the queue.

\subsubsection{Traffic load}
Traffic load is the extent to which a node participates in communication. In WRSNs, a node having high traffic consumes energy much faster than a leaf node since it has to receive, process and forward packets. Protecting the relay availability of high-load nodes is crucial for ensuring end-to-end data delivery. Base station estimates load through periodic report, which gets mapped into a load-related score.



\subsubsection{Estimated UAV travel time}
The estimated travel time of the UAV from its current location to a candidate node reflects the service cost in terms of delay and propulsion energy. Prolonged travel time causes the node to require more energy to reach and delays charging of other waiting nodes. The base station determines this estimate from the up-to-date UAV state and the node's location. When other attributes are similar, shorter travel times are preferred, but the scheduler specifically weighs urgency against mobility cost.


\subsubsection{Flight direction alignment}
The flight direction alignment determines whether a candidate node falls in the forward direction of the current UAV motion. When the scheduler selects nodes along the current heading, the UAV prevent frequent sharp turns or backtracking, thus limiting travel time and propulsion-energy expenditure.

The base station determines the alignment by using the angle between the UAV heading vector and the displacement vector to each node. When multiple candidates have similar urgency, the alignment score serves as a cost-sensitive tiebreaker.

cost-aware tie-breaker.

\subsubsection{Composite priority and queue maintenance}

The base station combines the four attribute scores into a single composite priority value for each node. The urgency score derived from residual energy and the traffic-load score both increase the priority, because nodes that are closer to depletion or carry more traffic should be served sooner. The estimated travel time decreases the priority,	because longer flights consume more energy and delay service to other nodes. The direction-alignment score increases the priority, because nodes that lie along the current flight path reduce the need for sharp turns and backtracking. A separate weight controls the relative influence of each factor, and all scores are normalized to a common range before combination.

The weight selection reflects the design rationale. The urgency weight is set high because preventing outages is the primary objective. The travel-time weight is set to discourage long detours.  The traffic-load and direction-alignment weights serve as secondary modifiers. Practitioners can tune these weights: increasing the	traffic-load weight protects relay-heavy topologies, increasing the direction-alignment weight favors compact trajectories. \textbf{Weight tuning guideline:} urgency weight 0.4, traffic-load weight 0.2, travel-time weight 0.25, and direction-alignment weight 0.15 (summing to one after normalization). Sensitivity is discussed in Section~V-E.

After computing composite priorities, the base station constructs the charging queue. The queue is updated online to reflect time-varying node states and UAV motion: a node is removed once it has been served,
and newly reported demands are inserted as they arrive. The base station recomputes priorities and reorders the queue whenever updated attributes alter the service order.

\textbf{Return-to-depot integration.}  The feasibility check is applied before a node is admitted into the active queue. Specifically, when evaluating whether to enqueue a new candidate, the base station first estimates the energy that would be required to fly from that candidate back to the base station using the current UAV position and the planned return path, and then verifies that the UAV's remaining energy is sufficient to reach the candidate, hover and charge there, and return safely. If not, the candidate is placed in a deferred list for the next mission. This ensures that the return-to-depot guarantee is not an afterthought but an integral part of the queue construction process.

\subsection{ISAC-Driven UAV State Estimation}

Online charging scheduling requires timely and accurate knowledge of the UAV state (position and velocity), since travel-time and direction-related priority terms depend on the current UAV motion. Although the UAV can provide onboard state reports, these may be degraded by sensor errors and reporting latency. ISAC introduces a
complementary, network-side sensing modality that can support UAV tracking using communication waveforms and sensing processing.

When the UAV enters a region with pending charging demands, the base station schedules an ISAC sensing round and selects a subset of sensor nodes with favorable geometry and reliable links. The selected nodes transmit predefined reference waveforms and, in a time-separated sensing interval, listen to the returned signals to capture echoes associated with the UAV.  Each node then extracts compact observables (e.g., delay- and Doppler-related features) via lightweight
correlation-based processing and forwards them to the base station.

At the base station, delay-related observables are mapped to range information, and Doppler-related observables provide information about radial motion. The base station uses geometric localization to estimate the UAV's position and refines the velocity estimate based on the temporal evolution of measurements from multiple nodes.  


The resulting ISAC-based estimates are fused with the UAV-reported state using a Kalman-type filter. In essence, the filter maintains a predicted UAV state based on a dynamic motion model, when a new ISAC-derived measurement arrives, the filter computes a Kalman gain that balances trust between the prediction and the measurement, producing a corrected state estimate. This fusion mitigates two kinds of errors: sensor noise and bias in the UAV's onboard inertial navigation, which can accumulate as drift over time and outliers or inaccuracies in individual ISAC measurements, which are smoothed by the filter's prediction step. The updated state is then fed back to the queue-maintenance procedure to refresh the travel-time and direction-compatibility terms, which may reorder composite priorities as the UAV progresses along its route.

\textbf{Overhead and scalability.}  Node-side correlation takes milliseconds. Observables are a few bytes per round. BS-side fusion is $O(n)$ with $n = 3$--$5$, negligible against seconds-scale travel. Queue reordering is $O(K \log K)$ and remains tractable for several hundred nodes. End-to-end latency is tens to hundreds of milliseconds, small compared with travel times. For large networks, ISAC uses a geometry-selected subset, updates are event-driven, and the area can be zoned. Centralized control suits networks up to a few hundred nodes, distributed extensions are left for future work.

\textbf{Robustness.} The Kalman filter handles moderate dynamics and degrades gracefully when measurement quality drops. Independent safeguards, the return-to-depot constraint and the urgency score, keep the queue functional even if estimates lag. Performance trends remain consistent across $100--500$ nodes, indicating graceful degradation.

\subsection{Partial Charging with Time Allocation}

We adopt a partial-charging policy in which the UAV distributes hover time among queued nodes rather than fully recharging one at a time. Full recharge maximizes per-node delivery but risks missing deadlines for others. Partial charging serves more nodes per mission, improves coverage, and reduces outage probability.

A risk of partial charging is repeated near-depletion cycles. We mitigate this by defining charging demand to push nodes \emph{above} a safety margin, not merely to the threshold. The steep urgency score also prevents indefinite postponement of near-critical nodes.

Under the adopted WPT abstraction, once the UAV is positioned within an effective charging region above a node, the delivered energy increases approximately in proportion to the time spent hovering.

To compute the time allocation, the base station first uses each node's residual energy to derive its charging demand. Both residual energy and traffic load contribute to the urgency/criticality weight: a node with low residual energy and high traffic load receives a higher weight than a node with the same residual energy but lower traffic demand, because its failure would disrupt more network traffic. Specifically, the base station computes an urgency weight for each node by adding its residual-energy urgency score to a scaled version of its traffic-load score, where a tunable parameter controls how strongly traffic load influences the result. This per-node sum is then divided by the corresponding total across all nodes currently in the charging queue, yielding a normalized urgency weight.  The charging time assigned to a node is finally obtained by multiplying its urgency weight by the total hover time budget available for the current mission.

\textbf{Reallocation timing.}  The aggregated charging time and per-node allocations are recomputed \emph{each time the charging queue changes}, that is, whenever a new node joins the queue, a node is served and removed, or ISAC updates trigger a priority reordering. In practice, this means that time splits are recomputed before each hover, not on a fixed periodic schedule. This event-driven reallocation ensures that the time allocation always reflects the most current urgency distribution and UAV energy budget.

When a new queue node joins, the base station recomputes aggregate time and per-node allocations to keep the partial-charging plan consistent.

\section{Performance Evaluation and Discussion} \label{sec:performance}

\subsection{Simulation Configuration}
To evaluate the effectiveness of the proposed framework, we consider a simulated WRSN environment in a square monitoring area with numerous sensor nodes. The nodes are randomly distributed to capture spatial heterogeneity. The base station is located in the central part of the area and operates as a data sink and UAV depot. The summery of simulation parameters is given in Table~\ref{tab:sim_params}.


\begin{table}[!t]
	\caption{Simulation parameters.}
	\label{tab:sim_params}
	\centering
	\renewcommand{\arraystretch}{1.1}
	\begin{tabular}{lc}
		\toprule
		\textbf{Parameter} & \textbf{Value} \\
		\midrule
		Number of sensors & 100--500 \\
		Sensor communication radius & 50\,m \\
		Sensor sensing radius & 25\,m \\
		Sensor battery capacity & 10\,J \\
		Sensor energy consumption rate & 0.01\,J/s \\
		(sensing + processing + communication) & \\
		Threshold of charging request & 30\% residual energy \\
		UAV speed & 20\,m/s \\
		UAV propulsion power (flight) & 150\,W \\
		UAV hovering power & 200\,W \\
		UAV initial energy & 500\,kJ \\
		UAV RF emission power (WPT) & 200\,W \\
		Effective WPT charging rate & 5\,W \\
		Monitoring area & $500 \times 500$\,m$^2$ \\
		\bottomrule
	\end{tabular}
\end{table}

The sensor battery capacity and energy consumption rate are selected to represent a low-power sensor model in the simulator. Under the assumed consumption rate of 0.01\,J/s, the nominal battery lifetime is approximately 1{,}000\,s in continuous active operation. The UAV is initialized with 500\,kJ of energy, which is sufficient to support multiple flight and hovering operations during a mission. The UAV propulsion and hovering powers are chosen to reflect typical values for small to medium multirotor platforms. The effective 5\,W WPT charging rate represents the received charging power after accounting for propagation and conversion losses.

Sensor nodes have identical battery capacities but may experience different traffic loads. Some nodes generate more data or participate in more forwarding, which leads to heterogeneous energy depletion across the network. Nodes consume energy for sensing, processing, and communication, and their residual energy decreases accordingly. When the residual energy of a node falls below the predefined threshold, it issues a charging request to the base station.

The UAV begins each mission at the base station with a fixed energy budget. Its speed and propulsion power are set to representative values for a small to medium multirotor UAV. Before the first charging decision is made, the UAV remains at the base station; once dispatched, it flies in a straight line to the first scheduled service location. Thereafter, the UAV trajectory is determined by the charging schedule.

The proposed ISAC-enabled framework (OD-UCS) is compared with two representative baselines. The first baseline is CGDA-Q~\cite{2.1}, which constructs charging cells to determine UAV hover locations, dynamically assigns charging tasks among multiple UAVs to balance load, and applies Q-learning for charging-path planning. The second baseline is MA-DDQN~\cite{2.3}, which jointly optimizes dynamic charging decisions and UAV trajectories to reduce task completion time and improve sensor survivability.

Both baselines are evaluated under identical network topologies, sensor placements, and traffic patterns. Although CGDA-Q was originally designed for multi-UAV operation, it is configured with a single UAV in this study to ensure a fair comparison under the same system scale.

\begin{figure*}[!tp]
	\centering
	\subfloat[Energy Usage Efficiency]{\includegraphics[width=.28\linewidth, height=3.2cm]{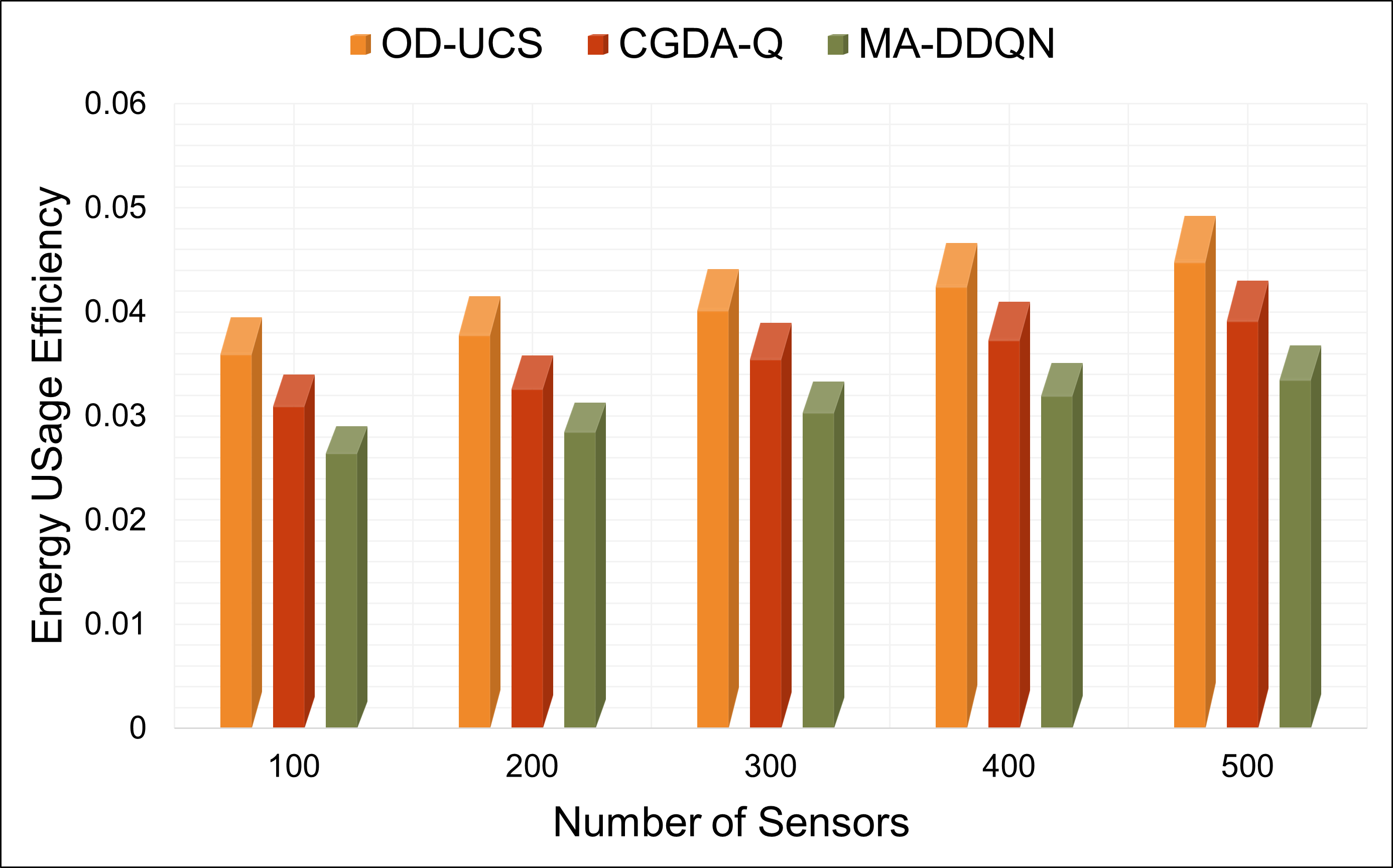}}\quad
	\subfloat[Travel Distance]{\includegraphics[width=.28\linewidth, height=3.2cm]{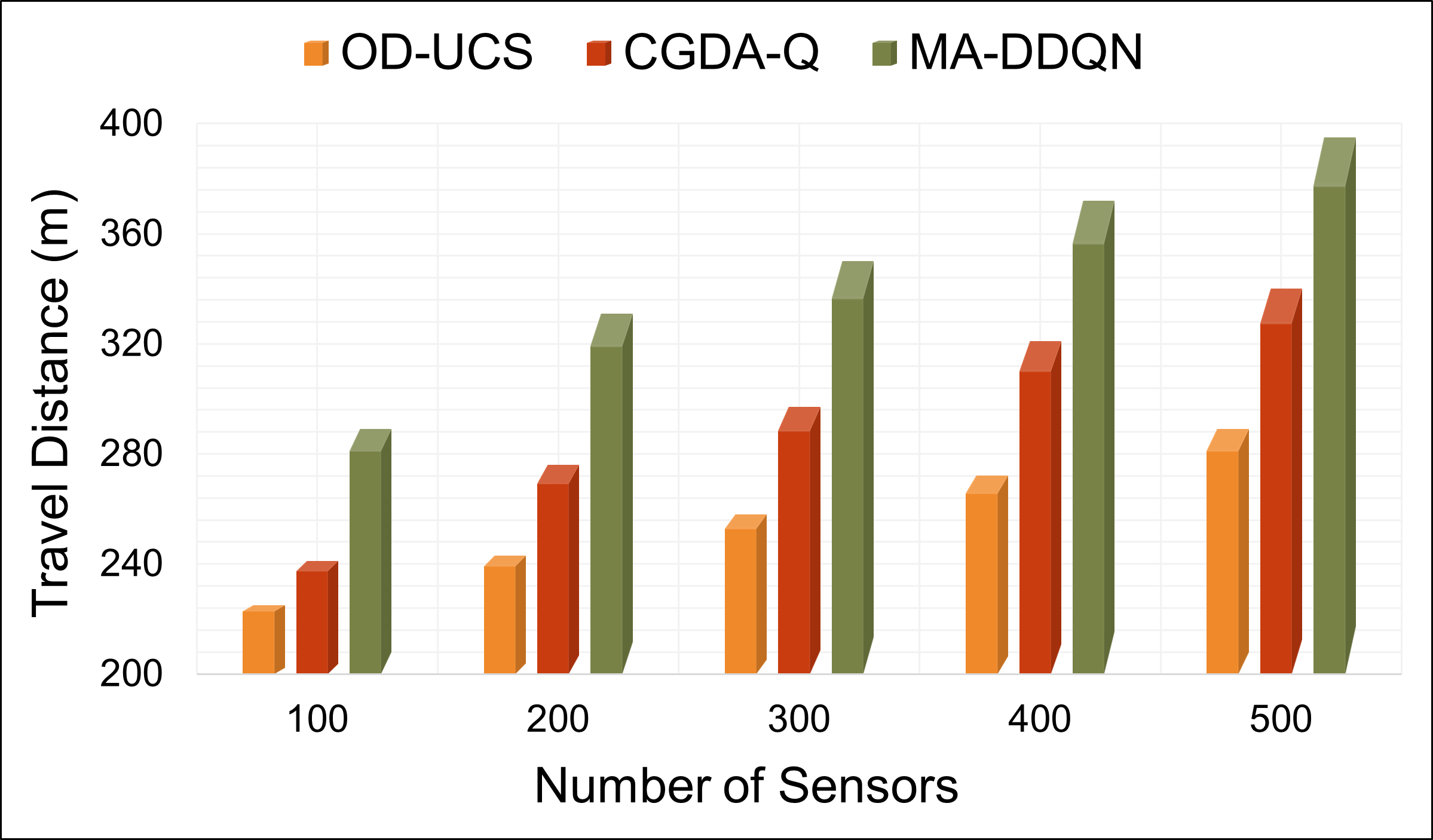}}\quad
	\subfloat[Charging Delay]{\includegraphics[width=.28\linewidth, height=3.2cm]{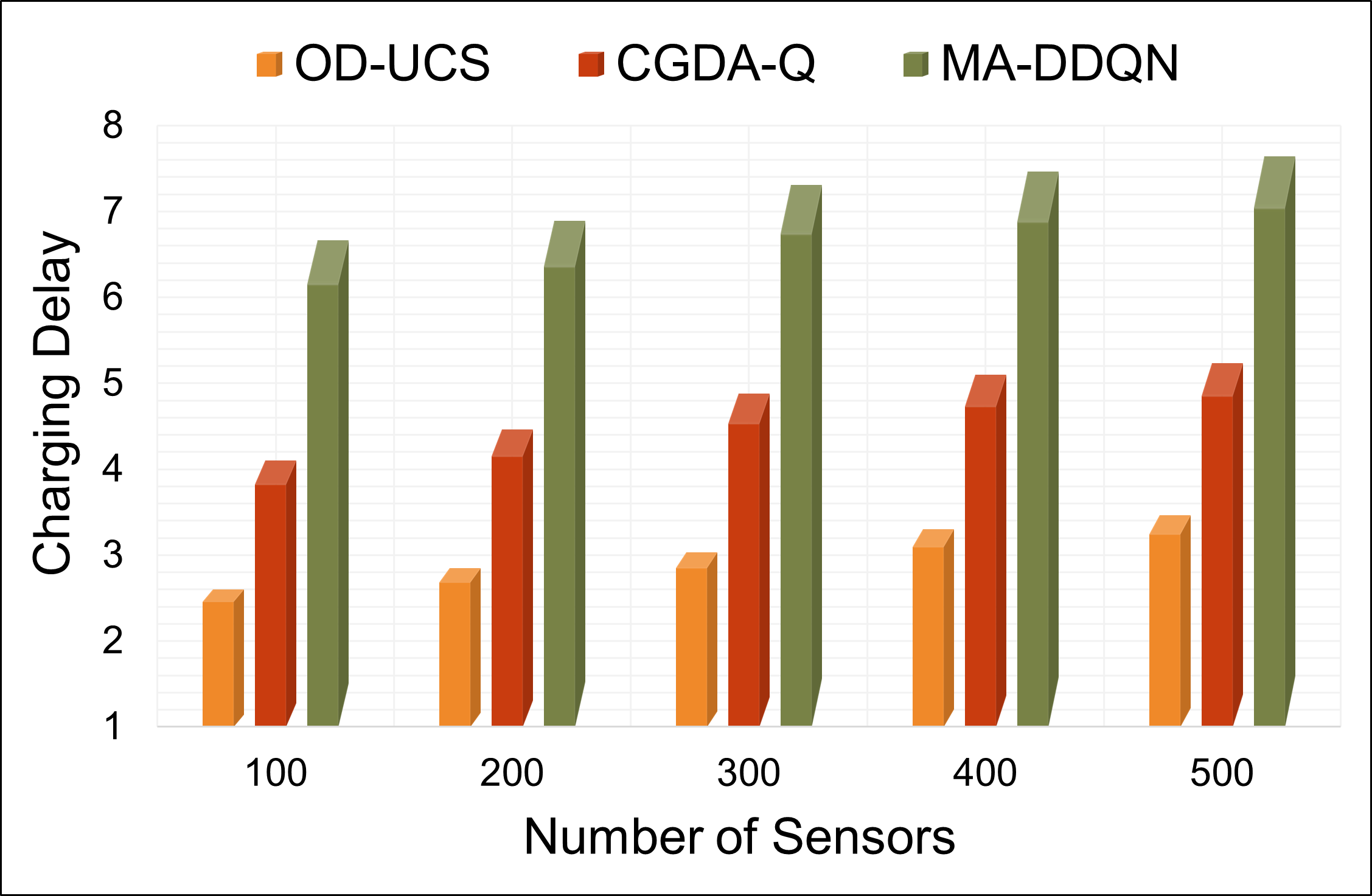}}
	\caption{Performance over sensor nodes.}
	\label{fig:Psn}
\end{figure*}

\subsection{Energy Usage Efficiency}

Fig.~\ref{fig:Psn}a shows that the proposed scheme achieves consistently higher energy usage efficiency than CGDA-Q and MA-DDQN across all node densities, meaning that a larger fraction of the UAV's onboard energy is converted into effective charging rather than propulsion and hovering overhead. This gain becomes more pronounced as the network becomes denser, indicating that the priority design scales gracefully.

\subsection{Travel Distance}

As shown in Fig.~\ref{fig:Psn}b, the proposed scheme achieves a shorter average UAV travel distance per tour than the two baselines, resulting in more compact and directionally coherent trajectories. The reduced path length directly reduces the propulsion energy consumption and provides more of the limited energy budget of the UAV to wireless power transfer.

\subsection{Charging Delay}

As shown in Fig.~\ref{fig:Psn}c, the proposed scheme reduces the average charging delay much more than CGDA-Q and MA-DDQN, and can serve more charging requests in the same operation time. This indicates that the policy not only improves the service speed of energy-critical nodes, but also the temporal fairness in general.

\subsection{Sensitivity to Priority Weights}

To assess the impact of weight selection on performance, we varied the urgency weight $\alpha$ from 0.2 to 0.6 (adjusting $\gamma$ inversely, with $\beta$ and $\delta$ held constant). The results show that energy usage efficiency varies by less than 5\%\ across this range, and the relative ranking against both baselines is preserved. This insensitivity indicates that the framework is robust to moderate weight perturbations. The largest performance drop occurs at very low $\alpha$ ($\le 0.2$), where urgency is under-weighted and critical nodes are deprioritized, this boundary case reinforces the design rationale for setting $\alpha$ relatively high.

\section{Conclusion and Future Directions} \label{sec:conclusion}

This article has presented an ISAC-enabled, on-demand UAV charging framework for wireless rechargeable sensor networks that tightly couples queue-based charging decisions, mobility-aware service costs, and ISAC-assisted UAV state awareness under time-varying traffic and energy conditions. By jointly considering node urgency (residual energy and traffic load), UAV mobility cost (travel time and flight-direction compatibility), and partial-charging control, the proposed design improves the utilization of the UAV's limited onboard energy and enhances the temporal reliability of energy replenishment. Simulation results demonstrate that the framework consistently attains higher energy usage efficiency, shorter and more compact UAV trajectories, and significantly reduced charging delay compared with representative
baseline schemes.

\textbf{Key design insights.} First, coupling ISAC state estimation with online priority updates improves scheduling quality compared with approaches that rely on static information. Second, partial charging with urgency-weighted allocation serves more nodes per mission without creating near-depletion cycles, provided a safety margin is built into the demand calculation. Third, integrating the return-to-depot constraint into queue construction guarantees safe tours by design.

\textbf{Open challenges and future directions.}
Several promising research directions emerge from this work:
\begin{itemize}
\item \textit{Multi-UAV coordination:} Distributed scheduling, conflict-free trajectory design, and energy-aware task allocation.
\item \textit{Realistic charging and flight models:} Detailed WPT models, flight-energy models, and operational constraints such as no-fly zones.
\item \textit{Scalability:} Distributed or edge-assisted architectures for networks exceeding several hundred nodes.
\item \textit{Heterogeneous UAV platforms:} Fleets with different endurance, speed, and charging capabilities.
\item \textit{Security:} Integrity of charging requests and ISAC measurements against spoofing or jamming.
\end{itemize}

\balance

	\bibliographystyle{IEEEtran}
\bibliography{sample-base}

\end{document}